\documentclass[aps,prl,twocolumn,superscriptaddress,groupedaddress]{revtex4} 
\usepackage{amsfonts}
\usepackage{amsmath}
\usepackage{amssymb}
\usepackage{graphicx,color}
\usepackage{dcolumn}
\usepackage{times}
\usepackage{amsmath}
\usepackage{color}
\usepackage{ulem}

\begin{document}

\title{Phase domain walls in weakly nonlinear deep water surface gravity waves}

\author{F. Tsitoura}
\affiliation{Dynamics Group, Hamburg University of Technology, 
21073 Hamburg, Germany}

\author{U. Gietz}
\affiliation{Department of Fluid Dynamics and Ship Theory, Hamburg University of Technology, 
21073 Hamburg, Germany}

\author{A. Chabchoub} 
 \affiliation{School of Civil Engineering, The University of Sydney, Sydney, NSW 2006, Australia}

\author{N. Hoffmann}  
\affiliation{Dynamics Group, Hamburg University of Technology, 
21073 Hamburg, Germany} 
\affiliation{Department of Mechanical Engineering, Imperial College London, London SW7 2AZ, United Kingdom}

\begin{abstract}

We report the theoretical derivation and the experimental as well as numerical observation of nonlinear phase domain walls in weakly nonlinear deep water surface gravity waves. The domain walls presented are connecting homogeneous zones of weakly nonlinear plane Stokes waves of identical amplitude and wave vector but differences in phase. By exploiting symmetry transformations within the framework of the nonlinear Schr\"odinger equation we demonstrate the existence of exact analytical solutions representing such domain walls in the weakly nonlinear limit. The walls are in general oblique to the direction of the wavevector and stationary in moving reference frames. Experimental and numerical studies confirm and visualize the findings. While domain walls are well known from many other fields in physics where strong nonlinearities are involved, e.g. in the case of dispersive shock waves, the present findings demonstrate that nonlinear domain walls do also exist in the weakly nonlinear regime of general systems exhibiting dispersive waves.

\end{abstract}

\maketitle
Domain walls are transition zones of finite width between neighbouring homogeneous domains. They are of fundamental importance in many fields of modern physics. In wave dynamics, domain walls connecting neighbouring wave fields differing in wave amplitude and phase, are also called wave jumps \cite{dw_ferom, dw_prl, dw_optic}.

The question of the existence of domain walls between finite amplitude surface gravity waves on deep water, i.e. between domains of deep water Stokes waves, has been investigated at least since the 1960s \cite{Witham1, Peregrine1, rob, rob_per}. The dispersive nature of water waves shows that shape conserving domain walls do not exist in the linear limit of infinitesimally small wave amplitudes. In the case of large amplitude waves with strong nonlinearities involved, the question is part of the field of dispersive shock waves \cite{disp1, disp2,Trillo}. However, already early on there was also speculation about the existence of weakly nonlinear domain walls in the small but non-zero amplitude limit, where nonlinearity might counteract dispersion and might lead to the existence of shape conserving wall or jump type behaviour. While similar considerations in the context of localisation have led to the discovery of solitons and breathers \cite{Nailalt,JohnNat} in weakly nonlinear waves, there were only few attempts to derive and study analogous shape conserving solutions for domain walls or wave jumps.

Only recently a remarkable series of experimental and theoretical studies has sparked new interest in the field \cite{Hend1, Hend2, Hend3}. By performing wave tank tests a wave jump between neighbouring domains of fully out of phase Stokes waves has been studied. To explain the observed wave state, the authors introduced coupled systems of nonlinear Schr\"odinger equations. 

The present study has been motivated by these results, but starts from a different, simpler conceptual perspective. We employ a 2D+1 (D stands for the spatial dimensions) focusing nonlinear Schr\"odinger equation (NLS) to obtain analytical solutions for weakly nonlinear domain walls. The NLS is the universal lowest order equation describing the spatio-temporal dynamics of weakly nonlinear narrow-banded wave packets \cite{Zakh, BenneyNewell}. The 2D+1 form has been used for various questions on stability and non-planar solutions \cite{BenneyRoskes, YueMei, PihlMei} and can be derived by the method of multiple scales \cite{Hasimoto, Mei, D_S},
\begin{eqnarray}
i  \left( \frac{\partial A}{\partial t} + c_g \frac{\partial A }{\partial x} \right) - 
\alpha \left(  \frac{\partial^2 A }{\partial x^2} -2 \frac{\partial^2 A }{\partial y^2} \right) 
- \beta |A|^2 A=0,
\label{nls1}
\end{eqnarray}
where $A(x,y,t)$ is the complex wave envelope, $t$ denotes time, $x$ and $y$ are the orthogonal horizontal spatial coordinates, with uni-directional background wave propagation along the $x$-axis, and $c_g$ is the group velocity. The dispersion and nonlinearity coefficients $\alpha$ and $\beta$ are functions of frequency $\omega$ and wavenumber
$\kappa$, and for deep water waves result as

\begin{eqnarray}
\alpha &=&  \frac{\omega}{8 \kappa ^2}, ~~~ \beta= \frac{\omega \kappa ^2}{2}. 
\end{eqnarray}

The linear dispersion relation is $\omega =\sqrt{g  \kappa}$, with $g$ the constant of gravity. To first order in wave steepness, the surface elevation $\eta \left( x,y,t \right)$ 
is given as  
\begin{eqnarray}
\eta\left(x,y,t\right) = {\rm Re} \Big \lbrace  A \left(x,y,t \right) \exp %
\big[ i\left( \kappa x -\omega t\right) \big] \Big\rbrace.  
\label{eta}
\end{eqnarray}

For simplicity, we consider a scaled form of the 2D+1 NLS,  
\begin{eqnarray}
iu_T + u_{XX} - 2 u_{YY} +2 |u|^2 u=0,
\label{nls}
\end{eqnarray}
which is obtained from Eq.~(\ref{nls1}) by introducing the scaled variables
\begin{align}
\label{scale}
\begin{split}
X &= \left( x-c_g t \right), ~ Y= y, \\ 
T&= -\alpha t, ~~ u= \sqrt{\frac{\beta}{2 \alpha}} A.  
\end{split}
\end{align}

A transformation introduced earlier \cite{hh}, based on the idea to investigate solutions of the 2D+1 NLS that actually depend on one spatial direction only, allows the derivation of solutions for the 2D+1 NLS from solutions of a 1D+1 NLS. For that purpose we introduce an angle $\gamma$ and a new spatial coordinate $Z$ according to 

\begin{eqnarray}
Z= \frac{X\cos \gamma + Y \sin \gamma}{\sqrt{|1-3 \sin^2 \gamma|}}. 
\label{Z}
\end{eqnarray}

We obtain 
\begin{eqnarray}
i \frac{\partial u}{\partial T} + s  
 \frac{\partial^2 u}{\partial Z^2} + 2 |u|^2 u =0, 
\label{u_Z}
\end{eqnarray}
which is indeed a 1D+1 NLS, involving a sign factor $s=\left( 1 -3 \sin^2 \left( \gamma \right)\right) / \left(|1-3\sin ^2 \left(\gamma \right)| \right)$.

A solution of the 1D+1 NLS Eq.~(\ref{u_Z}) thus corresponds to a solution of the 2D+1 NLS Eq.~(\ref{nls1}). Depending on $\gamma$, two cases result. For small values of $\gamma < 35.26^{o}$, $s=+1$ and the 1D+1 NLS is focusing. For $\gamma> 35.26^{o}$, $s=-1$ and the 1D+1 NLS turns out defocusing. While the former case has already been studied  \cite{yuen_lake, saffm}, we will consider the latter case here. 

For the present purpose of identifying the desired shape conserving stationary or quasi-stationary domain wall solutions in the 2D+1 focusing NLS, we choose the corresponding shape conserving spatially localised nonlinear solutions of the 1D+1 defocusing NLS: black and grey solitons \cite{Sh_Zak, Akhm_book, Chab_dark, Chab_gray, Fr_rev, lumps, lump_exp}, also referred to as dark solitons.

Grey and black solitons of the 1D+1 defocusing NLS possess exact analytical solutions of the form
\begin{eqnarray}
u\left(Z, T\right)=\!\!\!\!&\big [&\!\!\!\!\sin %
\left(\theta \right)\!+\!i \cos \left(\theta \right) 
\tanh\! \big \lbrace \!\cos \left(\theta \right) %
\!\big [ Z+2 \sin \left(\theta \right) T \big ]  %
\big \rbrace \big ] \nonumber \\
&\times& { \rm exp} \left(-2i T+ i \chi \right).
\label{uu}
\end{eqnarray}

Dark solitons of the 1D+1 defocusing NLS connect wave domains of identical amplitude which are out of phase. The black soliton, $\theta=0$, represents the case which goes along with a phase change of $\pi$ and drops of the wave envelope to zero, while for grey solitons, $\theta \neq 0$, the amplitude drops and the phase difference is less marked.

With this we can now construct domain wall solutions for the 2D+1 focusing NLS, or small amplitude weakly nonlinear deep water surface gravity Stokes waves. Indeed, we have two free parameters at hand. The first parameter is $\theta$, originating from the one-parameter family of grey/black solitons. The second parameter is $\gamma$, originating from the transformation between 1D and 2D formulations. The carrier wave amplitude $a$ and frequency $\omega$ may be understood as further parameters.

Combining all of the above, the solution of Eq.~(\ref{uu}) can be transformed, using the transformations ~(\ref{scale}), into a solution of Eq.~(\ref{nls1}). It may be written, with an additional constant $\chi$ as
\begin{eqnarray}
A\left(x, y, t \right)&=& -\frac{a}{\sqrt{2}\kappa^2} \big [ \sin \left(\theta \right)  + i \cos\left( \theta \right) 
\tanh \big \lbrace \cos \left(\theta \right) M \big \rbrace \big ] \nonumber \\
&\times& {\rm exp} \left( -2 i  a^2 \alpha t + i \chi  \right), 
\label{A}
\end{eqnarray}
where 
\begin{eqnarray}
M &=& a \Bigg[ 
\frac{x \cos \left( \gamma \right) + y \sin \left( \gamma \right) }
{\sqrt{|1-3 \sin ^2 \left( \gamma \right) |}} + \nonumber \\
&+& \Bigg( 2  \alpha  \sin \left(\theta \right) - c_g
\frac{\cos \left(\gamma \right)}{ \sqrt{|1-3 \sin ^2 \left( \gamma \right) |} }  \Bigg) a t \Bigg].
\label{M}
\end{eqnarray}

In the following we will discuss the key properties of the resulting solutions. For simplicity, and ease of illustration, we will compare the results with experimental realisations obtained in a water wave tank.

%
\begin{figure}[h]
\centering
\includegraphics[scale=0.2]{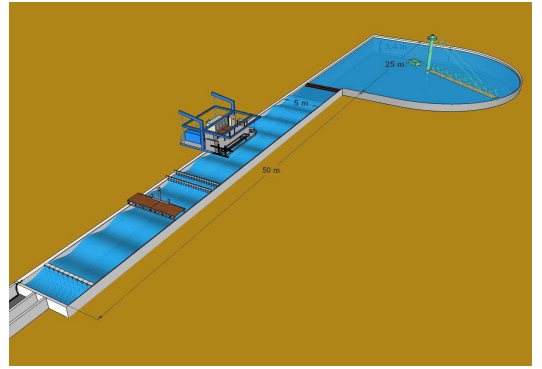}
\caption{(Color online) Schematic illustration of the water wave tank at the Hamburg ship model testing facility. Dimensions: $50 {\, \rm m} \times 5 {\,\rm m} \times 2.4 {\,\rm m}$.}
\label{fig1}
\end{figure}
%

The experiments have been conducted in a wave facility installed at the Hamburg ship model testing facility (HSVA). The wave tank is shown schematically in Fig.~\ref{fig1}. It has dimensions of $50~{\rm m}~\times~5~{\rm m}~\times~2.4~{\rm m}$. The wave maker consists of 10 flaps, each with a width of $0.5$ ${\rm m}$. The paddles are installed at one end of the tank and an absorbing beach of 10 ${\rm m}$ is located at the other side. The resulting propagation distance for the waves is about 40 ${\rm m}$. Lines of markers are installed at various distances from the wave maker. Each line has 25 markers where the first is centered at the middle of the tank and the others are equally spaced towards the tank wall. The motion of the markers is measured by a VICON (MX-3+) camera system with a sampling frequency of 50 ${\rm Hz}$. 

We choose an ($x,y$) coordinate system with $x$ along the tank and $y$ in the transverse direction.
The paddles are labeled by an integer $j\in [1,10]$, and in the following $y_j$ indicates the transverse
location of the center of the $j$-th paddle, assuming all of them located at 
$x=0$. The time-dependent displacement of each paddle can be written in the form of a time-dependent amplitude $a_j(t)$ and a time-dependent phase $\phi_j(t)$,
\begin{eqnarray}
\eta_{j} \left(a_j,y_j \right)={\rm Re} \Big \lbrace a_j \exp\big (i \phi_j \big )  \Big\rbrace.
\label{h_j}
\end{eqnarray}
The motion of the paddles is set according to the domain wall solutions described above. To evaluate the influence of the nonlinearity, all tests have been conducted for varying degrees of wave steepness.


First we discuss the domain wall arising from the black soliton, $\theta=0$, and a transformation angle of $\gamma = \pi /2$. The paddles move with a phase variation in the $y$-direction corresponding to a hyperbolic tangent.

\begin{figure}[pt!]
\centering
\includegraphics[scale=0.35]{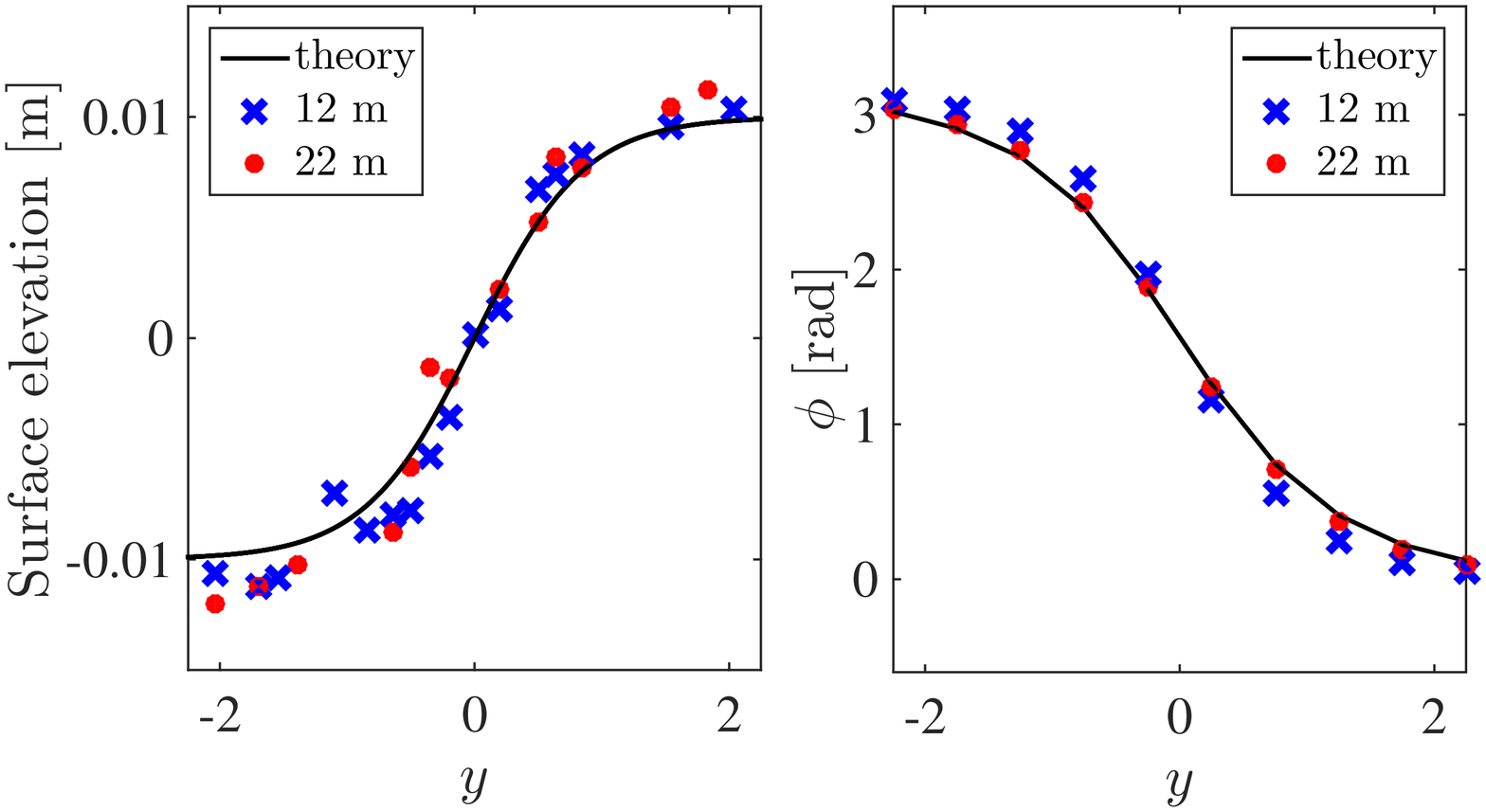}  \\
\includegraphics[scale=0.35]{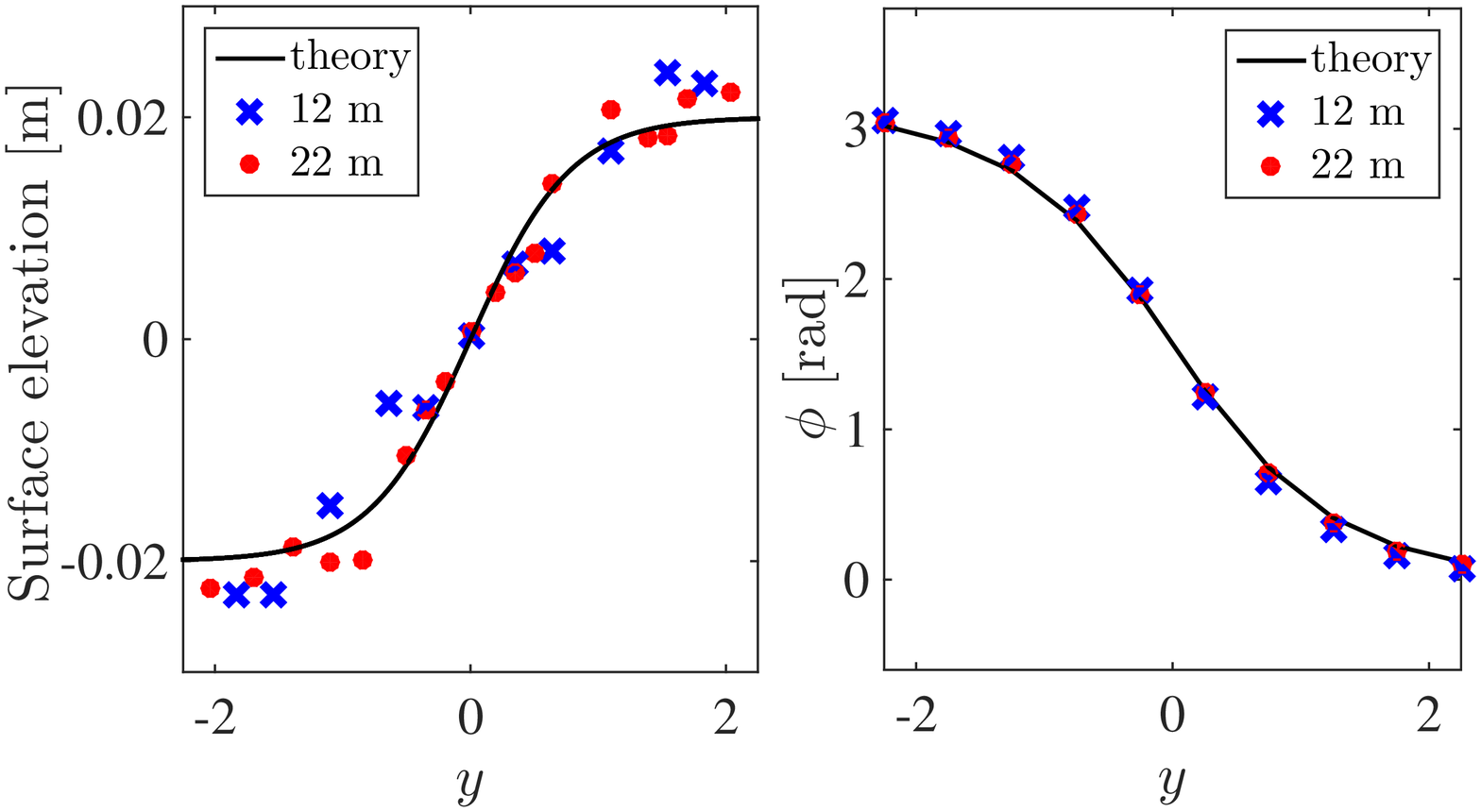}  \\
\includegraphics[scale=0.35]{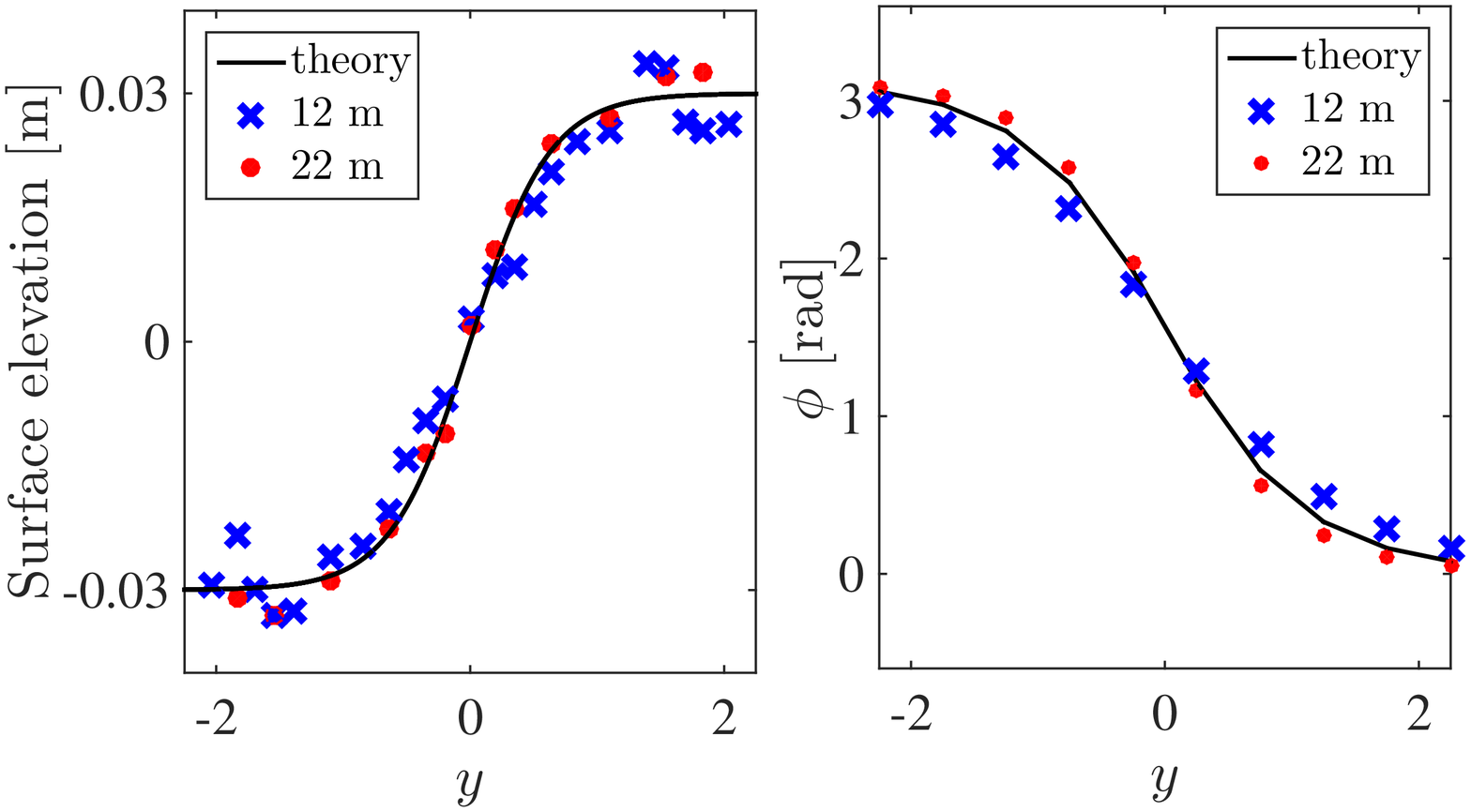}  
\caption{(Color online) Surface elevation and phase measured at a distance 
of 12 ${\rm m}$ and 22 ${\rm m}$ ((blue) crosses and (red) dots) from the wavemaker. 
$a \kappa = 0.11$, $0.16$ and $0.22$ (upper, middle and lower panels).
Solid (black) lines indicate the theoretical curves.
}
\label{fig2}
\end{figure}

Three different carrier wave steepness values, $a\kappa=0.11$, $a\kappa=0.16$ 
and $a\kappa=0.22$ have been tested. All of them are below the breaking threshold of Stokes waves \cite{babanin2012book}. The wavelengths, $\lambda = 0.58{\,\rm m}$, $0.78{\,\rm m}$ and $0.86{\,\rm m}$ have been selected to ensure the effects of surface tension to remain negligible, and the amplitudes of the background take the values  $a=0.01{\,\rm m}$, $0.02{\,\rm m}$, and $0.03{\,\rm m}$. Fig. \ref{fig2} shows the measurements of surface elevation and phase.

\begin{figure}[h]
\centering
\includegraphics[scale=0.12]{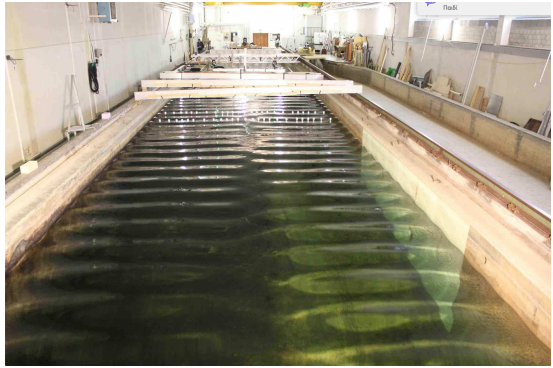}   
\caption{(Color online) Photographic image taken at a distance of 35 ${\rm m}$ from the wave maker 
for the experiment presented in the top panel of Fig.~\ref{fig2}: $a=0.01 {\,\rm m}$, $\kappa=10.8 {\,\rm m}^{-1}$.}
\label{fig3}
\end{figure}

It turns out that a stationary wall separating two wave domains of weakly nonlinear waves with identical amplitude and wave vector, however fully inverted phases results. The domain wall is aligned with the propagation direction of the underlying Stokes waves. A photography of one of the weakly nonlinear domain walls is shown in Fig.~\ref{fig3}. 

The measurements show very good agreement with the theoretical predictions from Eq. (\ref{A}): The domain walls stretch out along the whole length of the tank without any noticeable change in properties.

\begin{figure}[h]
\centering
\includegraphics[scale=0.3]{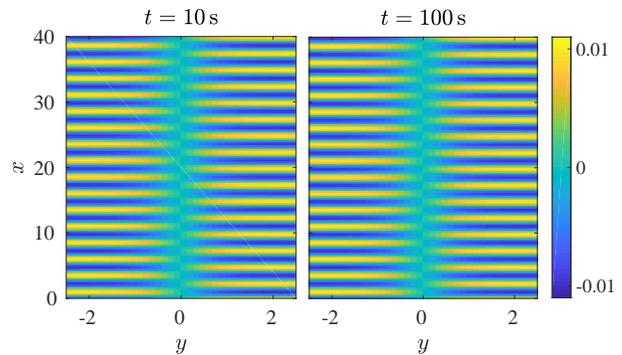}
\caption{(Color online) Simulation results for $u(x,y)$ at $t=10  {\,\rm s}$ and $100 {\,\rm s}$.
Parameters as in the top panel of Fig. \ref{fig2}: $a=0.01 {\,\rm m}$, $\kappa=10.8 {\,\rm m}^{-1}$. 
}
\label{fig4}
\end{figure}

We have also conducted some straightforward direct numerical simulations of the 2D+1 NLS equation~\eqref{nls} using a (second-order) finite difference scheme for the spatial discretization and a fourth-order Runge-Kutta method (with fixed time-step) for the time marching. Fig. \ref{fig4} shows the results 
for the reference case of lowest carrier wave steepness under study. Two different instants of time are depicted, $t=10$ ${\rm s}$ and $100$ ${\rm s}$. The numerical simulations suggest that for the time and length scales considered, the present domain wall pattern is not showing any sign of instability but remains stationary.%

%
We now investigate walls arising from general transformation angles $\gamma \neq \pi /2$. The initial conditions for the wave paddles are again determined by Eq.~\eqref{eta}. We investigate two cases: the first originates from the black soliton, $\theta=0$, and transformation angle $\gamma=0.4\pi$. The second one starts from a gray soliton, $\theta=\pi/6$, and a transformation angle $\gamma=0.35 \pi$.

\begin{figure}[h]
\centering
\includegraphics[scale=0.35]{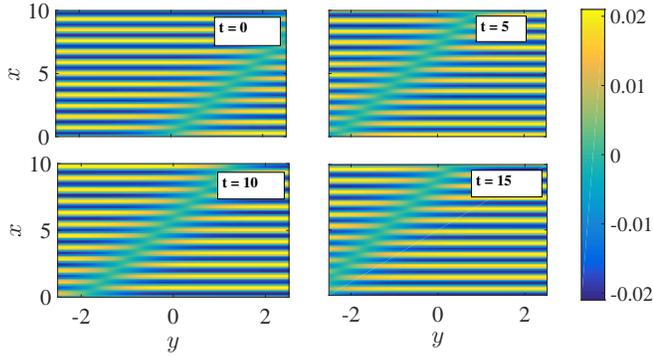}   
\caption{(Color online) Snapshots of the motion of a moving wall as resulting from Eq.~(\ref{eta}) for four different times $t=0$, $5$, $10$ and $15  {\,\rm s}$.  
$a=0.02 {\,\rm m}$, $\kappa=7.3 {\,\rm m^{-1}}$, $\chi=0$, $\theta=0$ and $\gamma=0.4 \pi$. }
\label{dif_x}
\end{figure}

\begin{figure}[h]
\centering
\includegraphics[scale=0.28]{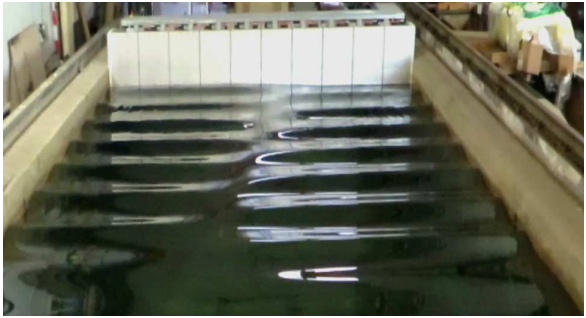}
\caption{(Color online) Photographic image taken at a distance of 10 ${\rm m}$ 
from the wave maker for $a=0.02 {\,\rm m}$, 
$\kappa=7.3 {\,\rm m^{-1}}$, $\chi=0$, $\theta=0$ and $\gamma=0.4 \pi$.
}
\label{gray1}
\end{figure}

Fig.~\ref{dif_x} shows the resulting analytical solution of the 2D+1 NLS for the first case. Snapshots of the surface elevation are shown for subsequent times. In contrast to the previous solution, where the domain wall was oriented in the direction of the wave vector, now an oblique boundary results, which moves laterally but is still stationary in a moving reference frame. This is the general result also for other choices of parameters, a photograph is given in Fig.~\ref{gray1}.

\begin{figure}[tp!]
\centering
\includegraphics[scale=0.35]{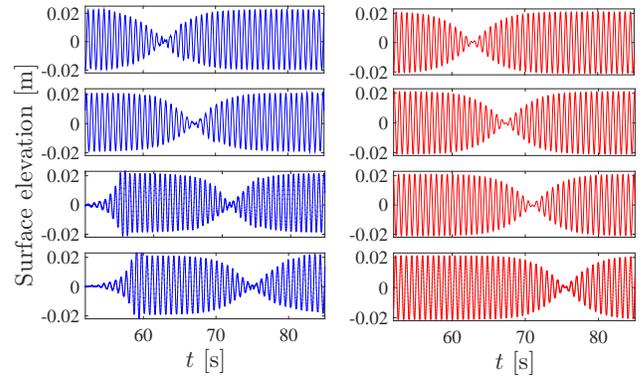}  
\caption{(Color online) Evolution of an oblique moving wall with 
$\theta=0$ and $\gamma=0.4 \pi$ for $a=0.02  {\,\rm m}$ and $\kappa=7.3 {\,\rm m^{-1}}$. 
Single point measurements (left) and corresponding theoretical predictions (right) at $x=3, 5, 8$ and 10 ${\rm m}$ from the wavemaker. 
}
\label{exp1}
\end{figure}

\begin{figure}[tp!]
\centering
\includegraphics[scale=0.35]{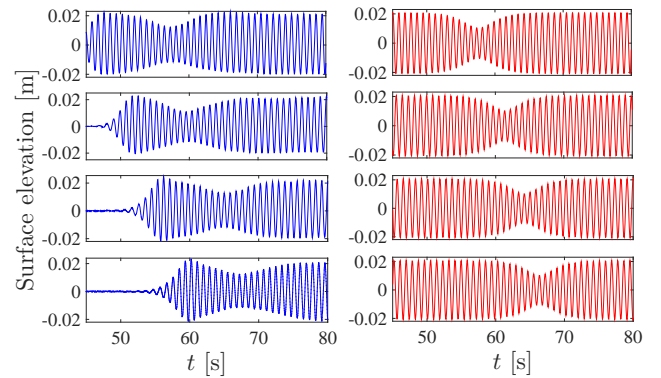}  
\caption{(Color online)
Evolution of an oblique moving wall with 
$\theta=\pi/6$ and $\gamma=0.35 \pi$ for $a=0.02 {\,\rm m}$ and $\kappa=7.3 {\, \rm m^{-1}}$. 
Single point measurements (left) and corresponding theoretical predictions (right) at $x=3, 5, 8$ and 10 ${\rm m}$ from the wavemaker. 
}
\label{exp2}
\end{figure}

For a quantitative comparison between the solutions of the 2D+1 NLS and the tank test, we took measurements along the wave tank at distances of $x=3$, $5$, $8$ and $10$ {$\textrm m$} from the wavemaker. The measurements are shown in the left panels of Figs.~\ref{exp1} and \ref{exp2} for $\theta=0$ and $\theta=\pi/6$. When the domain wall passes the sensor locations, the amplitude experiences a depression as expected and corresponding well to the solution of the 2D+1 NLS. The experimental results agree very well with the new domain wall solutions of the 2D+1 NLS.

While the first observed domain wall solutions are stationary and oriented in the direction of the wave vector of the neighbouring domains, the present oblique walls turn out to link neighbouring wave domains with a phase difference of less than $\pi$, which comes at the expense of obliqueness and lateral motion of the domain wall. Two videos of such laterally moving oblique domain walls can be found in the Supplemental Materials.



%

To summarize our findings, we have demonstrated the existence of novel elementary types of domain 
wall solutions in the 2D+1 NLS. The new solutions suggest that weakly nonlinear domain walls, separating neighbouring domains of weakly nonlinear waves with identical wave vector but differences in phase, do exist. For a phase difference of $\pi$ the domain walls are stationary and oriented along the wave vector of the neighbouring domains. For other phase differences, the domain walls are obliquely oriented to the wave vector and stationary in laterally moving reference frames. An experimental illustration for the solutions has been obtained for weakly nonlinear deep water surface gravity Stokes waves. For the nonlinearity range and the length and time scales considered, the measurements are in very good agreement with the predictions of the 2D+1 NLS theory.

Future work will be devoted to stability properties of the observed patterns. Further numerical studies will focus on limitations of the weakly nonlinear approach of the NLS in water waves. Moreover, similar studies in other universal fields of physics where weakly nonlinear dispersive waves arise, may be motivated from the present findings.

\section*{Acknowledgments}
The authors acknowledge support by Deutsche 
Forschungsgemeinschaft grant HO 3852/10. 


\end{document}